\documentclass{llncs}
\usepackage{lncsedoc}
\usepackage[final]{graphicx}
\begin{document}
\title{A Model of the Measurement Process in Quantum Theory}

\author{ Hans H. Diel}

\institute{Diel Software Beratung und Entwicklung, Seestr.102, 71067 Sindelfingen, Germany,
diel@netic.de}
\maketitle

\begin{abstract}
The so-called measurement problem of quantum theory (QT) is still lacking a satisfactory, or at least widely agreed upon, solution. A number of theories, known as interpretations of quantum theory, have been proposed and found differing acceptance among physicists. Most of the proposed theories try to explain what happens during a QT measurement using a modification of the declarative equations that define the possible results of a measurement of QT observables or by making assumptions outside the scope of falsifiable physics. This paper proposes a solution to the QT measurement problem in terms of a model of the process for the evolution of two QT systems that interact in a way that represents a measurement.
The model assumes that the  interactions between the measured QT object and the measurement apparatus are "normal" interactions which adhere to the laws of quantum field theory. This causes certain limitations associated with QT measurements. 
\end{abstract}


\section{Introduction}

QT is largely a theory of the possible results of measurements in subatomic physics. To a limited degree, QT also describes exactly what happens before a measurement takes place. Even less is known about how the multitude of alternative measurement results is converted to a definite result during a measurement. A number of alternative theories exist on the latter subject and are called interpretations of QT (e.g., Copenhagen interpretation, many worlds interpretation, or transactional interpretation, see \cite{Sexl}). The term "interpretation" gives the impression that these theories are not considered a necessary part of the QT but rather some optional ideas on top of QT. So far, there does not seem to be general agreement among physicists of the correct, or at least most plausible, interpretation of QT. 

Part of the "measurement problem of QT" and the quest for the right interpretation of QT is the search for a plausible theory for the "collapse of the wave function". A measurement is assumed to imply the collapse of a wave function, and the various interpretations of QT suggest details on when and how this collapse may occur. One of the interpretations, the many worlds interpretation, denies a collapse of the wave function and suggests an alternative to collapse. 
Thus, the collapse of a wave function (or some alternative explanation of why there is no collapse) is an unsolved problem in QT that is intimately connected to the measurement problem.

The author suspects that the reason for the unsatisfactory progress on the measurement problem is mainly because all of the suggested interpretations of QT (except for the transactional interpretation, see \cite{Cramer}) attempt to explain what is happening during a measurement using new ideas based on static declarative equations (such as the Schr\"odinger equation).
The author believes that the measurement problem is mainly a problem of understanding the \emph{process} involved in a QT measurement and that a satisfactory solution can only be found using a process-based description of QT in general, and QT measurement in particular. This process-based description of a theory is called a functional description (or functional interpretation, or functional model) by the author. 
The paper proposes a functional model of QT Measurement.
Starting with the assumptions that a measurement always contains at least one interaction between the measured QT object and the measurement apparatus, the model focuses on interactions in QT. 

The QT area that addresses the interactions between particles and between particles and fields is quantum field theory (QFT). QFT addresses interactions between particles to a great extent. QFT with its Feynman diagrams and description of scattering processes may also be viewed as containing some elements of a process-based description. Because interactions are the major constituent of the QT measurement process, the model had to be based on QFT. Assuming that the results of interactions are restricted to the capabilities defined by QFT, leads to a "natural" explanation for some of the limitations of QT measurements. 

The paper begins with a description of what the measurement problem is (section 2).
The proposed model is described in section 3. Finally, section 4 reviews the questions that have been associated with the measurement problem in the context of the proposed model of the QT measurement process.
 
\section{What is the QT Measurement Problem?}

A model of the measurement process in QT has also to address the measurement problem of QT. This model of the measurement process includes assumptions and theories of specific aspects that are usually associated with the measurement problem. 
\\
Wikipedia \cite{wikimeasure} writes on the measurement problem
\\
"The measurement problem in quantum mechanics is the unresolved problem of how (or if) wave function collapse occurs. The inability to observe this process directly has given rise to different interpretations of quantum mechanics, and poses a key set of questions that each interpretation must answer. "
\\
The key questions to be answered by an interpretation also apply to the proposed model of the measurement process. The set of key questions (or at least their phrasing) differs depending on certain basic assumptions used as a starting point. 
A well-founded description of the QT measurement problem is provided in  \cite{Maudlin1}.
Below some key questions addressed by the various interpretations of QT, including the model of the measurement process, are listed. 
To describe the measurement problem a very short overview on the relevant laws of QT is useful. 

The laws of QT say that the state of a quantum object can be described by a set of operators for various quantities, such as position X, momentum P, angular momentum S, etc. 
Differing from classical physics, the state of a quantity in QT (called an observable), e.g., momentum, cannot, in general, be expressed by a single definite value but requires multiple complex numbers called probability amplitudes. 
The probability amplitudes are assigned to possible values for the operator. A probability amplitude enables the computation of the probability that a measurement will result in that value. An operator is said to be in a state where it has a definite value, only if one of the probability amplitudes indicates a probability of 1 for the operator.

The laws of quantum physics define the dynamic evolution of the state of a quantum object and the interrelations between the states of the various operators representing the quantum object. These laws imply 
\begin{enumerate}  
\item Operator states with a definite value are rather exceptional. In general, multiple different values with probabilities $ 0 < probability(value_{i}) < 1 $  are possible.
\item The relationships between operators are such that specific pairs of operators (for example, position X and momentum P) cannot have concurrently definite values.
\end{enumerate} 
In the following text, the major subjects related to the measurement problem are described. Each section ends with questions that are unresolved or at least controversial. 

\subsection{Reduction} 
The root of the measurement problem lies in the fact that measurements can always show only definite values. If it were possible to extract the complete state with all associated probability amplitudes using a single measurement of a quantum object (including those for non-definite values), there would not be much left of the so-called measurement problem.
As a consequence of the rule that the measurement can only deliver definite values, the measurement has to include a reduction of the multiple non-definite values to a single definite value. When the multiple probability amplitudes describing the state of an operator are viewed as 
vectors in a vector space, the reduction may be viewed as a projection of that vector to some base (plane). The base to which the projection applies is determined by the measurement apparatus.
The reduction (or projection) is not just part of the generation of a measurement value but modifies the wave function, which can be concluded from the fact that a repeated measurement with the same base results in the same value with certainty.
\\
Questions:
\begin{itemize}
\item Does measurement determine the state of a QT object?
\item Is the wave function complete?
\item Does measurement always have determinate outcome?
\item Why is it impossible to measure the non-definite (probability amplitude) values?
\item Why can measurements only communicate projected values?
\item Why can certain observables not be measured concurrently?
\item When (in the sequence of process steps or under which circumstances) does the transition from probabilities to facts occur? 
\end{itemize}

\subsection{Collapse of the wave function} 
A second major problem related to (and sometimes equated with) the measurement problem is a satisfactory theory with respect to the "collapse of the wave function". Assuming that the state of a quantum object, including its dynamic evolution, is described by a wave function, a collapse of this wave function has to be assumed when a measurement takes place.
The normal progression of a wave function, e.g., as described by the Schr\"odinger equation, describes a continuous linear evolution of the multiple probability amplitudes that represent the alternative measurement results. 
QT considers the multiple alternative paths to be in a superposition. A measurement enforces the selection of a single alternative as the measurement result, which may be viewed as the collapse of the overall wave function that is composed of the superposition of the multiple alternatives.

The many worlds interpretation of QT ( see \cite{Everett} ) denies the collapse of the wave function. This theory assumes that all of the alternative paths for possible measurement results continue to exist; however, each alternative is in a different world (or universe). For reasons described below (section 3.1), the model of the measurement process assumes a collapse of the wave function during the measurement process.
\\
Questions:  
\begin{itemize}
\item Does the wave function evolve according to the Schr\"odinger equation?
\item What are the criteria for the distinction between interactions that include a collapse of the wave function from those that do not?
\item Is a measurement always coupled with the collapse of a wave function?
\item Does the principle of parsimony favor no-collapse theories ?
\end{itemize} 

\subsection{Transition from Probabilities to Facts} 

QT consists mainly of the principles, rules, and equations that describe how the probabilities (in the form of probability amplitudes) dynamically evolve in various situations to enable the prediction of the probability of different measurement results. Thus, the ultimate transition to facts due to a measurement is an essential element of QT, although the theory does not say much about when (under which circumstances) and how this transition occurs. Quantum physicists do not seem to consider this lack of explanation as a deficiency of QT, except for possibly agreeing that the measurement problem exists.
\\
Questions:  
\begin{itemize}
\item 	Is the transition from probabilities to facts related exclusively to measurements?
\item Is the measurement process random or deterministic?
\item What type of interaction constitutes a measurement?
\item When (in the sequence of process steps and under which circumstances) does the transition from probabilities to facts occur?
\end{itemize}  

\subsection{Entanglement} 
The problems physicists have with explaining and understanding entanglement are usually not linked to the measurement problem. In fact, the entanglement problem (if there is any) is not a problem of the theory (the theory correctly predicts the unbelievable behavior of entangled particles) but the problem of finding a plausible dynamical evolution model, which is in agreement with the equations of QT and does not violate principles, such as locality and causality.
Because the effect of entanglement occurs with measurements, the subject is associated with the measurement problem. 
\\
Questions:  
\begin{itemize}
\item When (under which circumstances) is entanglement terminated ?
\item Does the termination of entanglement always imply modifications of the entangled quantum objects?
\item Does the modification of an entangled quantum object always terminate the entanglement?
\item Is it possible to have a hierarchy of entanglements?
\item Is entanglement always combined with superpositions?
\item Does the measurement process violate locality?
\end{itemize} 

\subsection{Lack of a Functional Description} 
A further problem, which is not exclusively related to measurement (and which is usually not mentioned as part of the measurement problem), is that there does not exist a theory of what exactly happens during the evolution of a wave function in an interaction. 
R. Feynman writes in his introductory book on quantum electrodynamics  (QED)  \cite{Feynman} : 
\\
"I have pointed out these things because the more
you see how strangely Nature behaves, the harder it is
to make a model that explains how even the simplest
phenomena actually work. So theoretical physics has
given up on that."
\\
In \cite{Diel1} a model of "how things function" is called a functional description. Thus, the claimed problem is that no functional interpretation of QT exists.

There are most likely many physicists to whom the lack of a functional interpretation of QT does not present a problem at all, and others who may feel that a functional description of QT would be of value, but its absence is just a minor drawback of the present state of physics. Looking at QT in general, the author leaves the judgment of the value of a functional description to the society of physicists. However, for the resolution of QT measurement problem, the author sees an urgent need for a functional interpretation of QT interactions. Items such as the "collapse of the wave function" and "transition from probabilities to facts" are elements of the wave function evolution process and can only be reasonably discussed if the entire evolution of QT interactions is understood or at least discussed.

Although QFT, the area of QT that addresses interactions in QT, contains steps towards a process-oriented description, it does not allow construction of a model that shows the evolution of wave functions (the intermediate states) during an interaction between particles.
\\
Questions:  
\begin{itemize}
\item Does this model of the measurement process provide possible solutions to the measurement
problem?
\item To which extent can this  model of the measurement process be verified?
\end{itemize}

\section{Model of QT Measurement Interactions}

\subsection{Basic Assumptions}

The following basic assumptions are made for the model of the measurement process:
\begin{enumerate}
\item Objective reality of the wave function
\\
Whether an element of a theory of physics may be considered to represent reality is a difficult philosophical question. For the present paper, the assumption of the objective reality of the wave function simply means that, for the model described, it is reasonable to assume wave functions as the major objects that determine the dynamic evolution of the QT system.
\item The measurement process includes a collapse of the wave function. 
\\
This assumption is made mainly for pragmatic reasons. If the model did not assume a collapse of the wave function, another function, such as a "branch into multiple worlds", would have to be assumed. The embedding of an alternative function into the overall process structure would not be less difficult than the case for the collapse function.
\item A measurement always implies interactions between the measured quantum object and the measurement environment.
\\
Measurements of QT observables can be performed using a variety of measurement devices, apparatus, and processes. All such measurement processes have to include at least one interaction where the measured object exchanges information with some other entity belonging to the measurement apparatus. 
\item The model of the measurement process is based on QFT
\\
QFT is the area of QT that addresses interactions. It would not make sense to construct a model of the measurement process  that is incompatible with QFT. The following major points are derived from QFT and affect the model:
\begin{itemize}
\item The set of particles leaving an interaction need not be the same as those starting the interaction.
\item Even when the particle types that leave the interaction are the same as those starting the interaction, they are not necessarily the same particles.
As a consequence, it may not be possible to identify the measured particle after the interaction.

\item In general, the result of an interaction consists of a multitude of alternative paths. A further interaction may be required to reduce this multitude to a definite result.
\item In general, the particles leaving an interaction are correlated or even entangled.
\end{itemize} 
\end{enumerate}
Assumptions (1) and (2) mean that the model of the measurement process may be considered representative of an objective collapse theory (see \cite{objcollt} on objective collapse theories).

Some terminology used in the following text has to be introduced. 
The model does not distinguish between a particle and the (associated) wave. Therefore, the term "particle/wave" will be used throughout the remaining text.
\footnote{In the literature on QT, some authors used the name "wavicles" for what here is called "particle/wave".}
Interactions are described as occurring between QT objects. QT objects can be particles (i.e., particle/waves), fields, and aggregate QT objects, such as nuclei or atoms. In this paper, interactions are discussed in the context of measurements, which means the interaction is performed between a "measured QT object" and another QT object that is part of the measurement apparatus. The second QT object is abbreviated as "MA-object". 

\subsection{What can be measured ? (States and Observables) }

Before we can discuss the measurement process, it is necessary to look at the items that are candidates for being measured. From the point of view of a physicist, the components of the state of a physical object are of interest. The laws and equations of the physics theories refer to entities that constitute the state of a system. In QT, these state components are called observables. 
Disregarding what may be considered the "real" nature of the observable, it would be sufficient and desirable for the physicist if the measurement of the state components (i.e., the observables) would deliver quantities (e.g., numbers) that can be used within the formulas of physics theories. 
\footnote{In reality, it works the other way around: The entities used in the formulation of physics theories are chosen with the goal that they are measurable.}

Within QT, the overall value of an observable (prior to a measurement) is given by a set of values (e.g., in form of vectors or matrixes). A further important entity within QT is the probability amplitude. QT assigns a probability amplitude (a complex number) to the possible value combinations of the observables. The sum of all of the probability amplitudes may be viewed as the "wave function". 
The assumption of the "objective reality of the wave function" (see previous section) means the assumption of the objective reality of the probability amplitudes (or some representation of them). Based on this assumption, measurements of the probability amplitudes would be of interest as well. A measurement always implies a transition from a probability to a fact, and within QT, this transition always implies a reduction or modification of the "facts before" (including the probability amplitudes) to the "facts afterwards", making such an explicit measurement of probability amplitude impossible or useless. 

The discussion of the state of a quantum system has to include the property of QT systems that the wave function, in general, consists of the superposition of multiple states, with each state having an associated probability amplitude. When R. Feynman invented his theory on quantum electrodynamics (QED), he viewed the alternative states as a superposition of their trajectories and called them paths  (see \cite{Feynman2}). The term "path" will also be used in this paper to refer to these entities.
\begin{verbatim}
system :={ pw, ....};  // set of particle/waves
pw :=type, {path, ...}; // type + set of paths
path:={ state-component,...};//statecomponents
state-component :=position OR momentum
                      OR spin OR ...;

system.state := {pw.state, ....};
pw.state := {path.state, ...};  // paths  
path.state:={state-component.state, ...},
                    amplitude; 
state-component.state:=value, ...}; // values
\end{verbatim}
An ideal measurement would include the ability to measure not only the values belonging to a single path but also all of the paths before the paths are destroyed or modified. The assumption of a wave function collapse, which is implied with a (measurement) interaction, prevents these ideal measurements.
\footnote{Alternative theories, which do not assume a collapse of the wave function, such as the many worlds theory, nevertheless lead to the same effect.}

In addition to the impossibility of measuring probability amplitudes and multiple paths of the wave function, there exist further limitations on the state components that can be measured or can be measured concurrently. These limitations were mentioned in section 2 as part of the measurement problem. The author claims that these limitations, which appear like general QT principles, are consequences of the fact that measurements require interactions and that interactions according to the laws of QFT are limited in their ability to exchange information between the interacting QT objects.

\subsection{Major Actions within the Interaction Process}

For the model of the measurement process, a number of key process steps, called actions, can be identified:
\begin{itemize}
\item Position determination (of interaction)
\item Transition from probabilities to facts
\item Determination of output particle types
\item Information exchange 
\item Generation of output paths and establishing new entanglements
\item Collapse of the wave function
\end{itemize}
In a more detailed functional model of QT interactions (see \cite{Diel2}), the actions will not be performed exactly in the order listed above, and the actions are partly interwoven.

Besides interactions which include the above-listed actions, there are also interactions in QT that do not cause a reduction to a definite value nor a collapse of the wave function. For example, when a photon interacts with (atoms of) a mirror, there is, in general, no reduction of the wave function, nor is there an information exchange between the photon and the atoms of the mirror.  Therefore, such interactions are not suited for measurements. 
Within this paper, only interactions which are suited for measurements and which imply the above actions are addressed.

\subsubsection{Position determination (of an interaction)}
QT treats the position of a particle as an operator like any other (except that it has a continuous spectrum). Within the model of the measurement process, the position has a special role insofar as an interaction always starts with a definite position  $ x_{int} $. 
\footnote{This statement is not exactly in accordance with QFT. In QFT, the scattering matrix is computed using the superposition of (i.e., the integral over) all possible interaction positions if the computation is performed for position space.}
Thus, if the wave function of the measured QT object consists of multiple paths with differing positions, the determination of the interaction position  $ x_{int} $ implies a reduction of the paths and, thus, a first step in the transition from probabilities to facts.

Within standard QFT, the treatment of an interaction involves the consideration of virtual particles
that are exchanged between the interacting real particles. Although the proposed model of QT measurement must maintain the effects of virtual particle exchange, it uses a modified concept of particle/wave fluctuation (pw-fluctuation).
A pw-fluctuation can be thought of as a temporary concentration and amplification of one or several particles/waves at a certain point in space. The following assumptions are essential in considering pw-fluctuations and their role in the model of the measurement process:
\begin{itemize}
\item Only one pw-fluctuation may be active at a given point in time for a particle/wave.
\item The position where the pw-fluctuation occurs can be anywhere within the space occupied by the involved particles/waves. The position is determined randomly as a function of the amplitudes of the involved particles/waves. 
\item A pw-fluctuation refers to a certain force type (electro-weak, strong, or gravity). This restriction means that only particle/waves supporting a common force type can share a pw-fluctuation and can interact.
\end{itemize}

The immediate effect of a pw-fluctuation is the temporary formation of an entity called an interaction-object (pw-ia-object). A pw-ia-object merges the information from the particles/waves that caused the pw-fluctuation. When the temporary pw-ia-object disappears again, the original particles/waves may persist (or may be reinstalled) or a different set of particles/waves may appear. Accordingly, the long-term effect of a pw-fluctuation can be one of the following:
\begin{enumerate}
\item nothing durable (this result may be the case for the majority of pw-fluctuations),
\item an interaction \emph{with} a collapse of the wave function,
\item an interaction \emph{without} a collapse of the wave function,
\item a particle decay (not further addressed in this paper).
\end{enumerate} 
The model of the measurement process assumes that, first of all, the position of the fluctuation is determined by considering the sum of the (positional) wave functions of all of the particles/waves involved. The fluctuation position is determined randomly. Next, a second particle is determined randomly out of the particles that have a non-zero amplitude for the fluctuation position. If the particles/waves consist of multiple paths, the affected paths are determined.
If the fluctuation is supported by a second particle/wave (i.e., if another particle/wave has pw-space points associated with the fluctuation position), an interaction between the two particles/waves occurs. If there is no second particle involved, the fluctuation applies to the single particle. 

\subsubsection{Transition from Probabilities to Facts}
As described above, the first step in the transition from probabilities to facts is already performed when the interaction position is determined. Further reductions of the wave function paths to a single path (in addition to position selection) are assumed to occur at the beginning of an interaction.
In addition to the transition from probabilities to facts due to path selection, there are further decision points within the  model that represent a transition from probabilities to facts. For example, the model assumes that the determination of the output particle types happens during the interaction process (see below, "Determination of output particle types").
\footnote{It is not clear (to the author) that this process is in agreement with QFT. However, it would be difficult to test such a possible deviation from QFT.}

The model of the measurement process assumes that the elimination of non-selected paths affects not just paths from the measured (i.e., interacting) QT object but also paths from possibly entangled QT objects. To achieve this, the model states that entangled particles have common paths, each path representing a possible measurement result.

\subsubsection{Determination of output particle types}
In general, QFT predicts non-zero probabilities for different combinations of exit particle types. The combination with identical entry and exit particle types is just one of the possible interaction results. 
For example, electron - positron scattering (i.e., Bhabha scattering) may result in an electron and a positron, in two tauons or in two muons. For the subject of this paper, interactions with changing particle types are not further addressed because they will most likely occur only with measurements of this specific subject. 

The rules that define which output particle types may result from specific entry particle types are defined by QFT and described in textbooks on QFT, such as  \cite{Dyson,Ryder,Griffiths,Mandl}).
The inclusion of this action  within the interaction function represents another instance of transition from probabilities to facts.
\footnote{It is not clear (to the author) whether this transition is exactly in accordance with QFT. } 

\subsubsection{Information exchange}

The exchange of information between the measured QT object (e.g., particle/wave) and the MA-object that belongs to the measurement apparatus is the most important action within a measurement process. The information exchange is prepared by the formation of the temporary pw-ia-object mentioned under "Position determination". Ideally, this action would communicate the exact state of the measured quantum object to the measurement apparatus. Two types of limitations, however, make it impossible to communicate the exact state:
\begin{enumerate}
\item The reduction to a single path (see above), which, according to the  model, happens at the beginning of the interaction, eliminates part of the state of the measured QT object.
\item The information about the state of the measured QT object can only be communicated to the QT objects that exit the interaction in terms of the modifications performed to the exiting QT objects. These modifications are determined by the laws of QFT. It cannot be expected that the respective laws of QFT support arbitrary information exchange. In contrast, the limitations on the type of information that can be exchanged (i.e., measured) can be derived from the laws of QFT. 
\end{enumerate}
The information exchange between the measured QT object and the MA-object is represented by the modifications to the MA-object as a function of the state of the measured QT object.
QFT enables the computation of the possible outcomes of the interaction in the form of a scattering matrix. For example, the equation for the scattering matrix element for the electron - positron interaction (i.e., Bhabha scattering) is
$$
\begin{array}{llll}
\displaystyle
   M =  (-ie)^{2} \bar{v}(\vec{p}_{2}, s_{2} )  \gamma_{\mu}  u(\vec{p}_{1}, s_{1} ) \\[+8pt]  
\displaystyle
(-ig^{\mu\nu}/(p_{1} + p_{2})^{2})
 \bar{u}(\vec{p}'_{1}, s'_{1} ) \gamma_{\nu}
  v(\vec{p}'_{2}, s'_{2} ) \\[+8pt] 
\displaystyle
 -  (-ie)^{2} \bar{u}(\vec{p}'_{1}, s'_{1} ) \gamma_{\mu}  u(\vec{p}_{1}, s_{1} )  \\[+8pt] 
\displaystyle
( -ig^{\mu\nu}/(p_{1}-p'_{1})^{2})
 \bar{v}(\vec{p}_{2}, s_{2} ) \gamma_{\nu}  v(\vec{p}'_{2}, s'_{2} ) 
\end{array}
\eqno \mbox{(1)}
$$
According to the usual QFT notation $ u() $  represents the entry electron, $ \bar{u}()$  the exit electron, $ \bar{v}$  the entry positron, and $ v() $ the exit positron. M is the probability amplitude. 
Equations which are similar to equation (1) exist for all kinds of interactions in QFT. 
As a function of definite states of the entry particles and exit particles the equations deliver a probability amplitude.
Given a specific state of the entry particles, a non-zero probability amplitude is obtained for a large variety of exit particle states. 
Thus, in general, an  interaction results in a distribution of probability amplitudes for ranges of states of the exit particles/waves. 

An ideal measurement would provide a bijective (i.e., unique) mapping of the state of the measured QT object  to the state of one of the exiting QT objects. However, as a consequence of the laws of QFT, as reflected in equations like equation (1), 
such a bijective mapping is not supported.

Instead, QFT interactions provide only what the author calls “probabilistic projections”. The term “projection” here expresses that the measurement result may be understood as a projection of the measured observable to the state of the MA-object. The term “probabilistic” refers to the property that the interaction result contains a range of states with differing associated probabilities, i.e., a probability distribution of states.
A more detailed analysis of equation (1) and similar ones for other kinds of QFT scattering, including computer simulations, performed by the author showed the following pattern for probabilistic projections with QFT interactions:
\begin{itemize}
\item The result of a probabilistic projection is a probability distribution for paths of possible measurement results.
(More precisely, probability amplitudes are associated with paths.)
\item The possibilities for achieving a bijective mapping (after the reduction to a single path) vary with the type of observable to be measured. Measurements of position, i.e. mapping of the interaction position to some state of the measurement apparatus is relatively easy. Spin measurement is more difficult. The easiest way to measure the spin is to have the measured QT object (e.g. particle) interact with a magnetic field. 
\item Nearly definite measurement values can only be reached in terms of peaks within the probability distributions.
\item The laws of QFT support ways to enforce peaks in the probability distribution. However, there does not seem to exist general rules for the kind of interactions which produce peaks in the probability distribution.
\end{itemize}
The computer simulations indicate (they do not yet prove) that peaks within the probability distributions can be augmented by
\begin{itemize}
\item an increased asymmetry between the measured QT object and the interaction object belonging to the measurement apparatus. Major candidates for increased asymmetry are
\begin{enumerate}
\item asymmetry with respect to energy, including masses,
\item particles interacting with fields,
\item particles interacting with bound systems (e.g. atom, nucleus, hadron).
\end{enumerate} 
\item repeated interactions of the same kind.
\end{itemize} 
Because of the lack of general rules for the enforcement of definite measurement results, it is the responsibility of the physicist to design the measurement apparatus in such a way that interactions with probabilistic projections with peaks in the probability distribution are performed. For example, “sharp” probabilistic projection of a particle’s spin to its direction of momentum are possible using an inhomogeneous magnetic field. A momentum direction can be easily mapped to a position. With the Stern-Gerlach experiment, both cases are combined to measure the particle’s helicity (ie., the projection of its spin).

\subsubsection{Generation of output paths and establishing new entanglements}

As described in the preceding section, QFT predicts that an interaction, in general, results in a multitude of possible output states and that each state has an associated probability amplitude. The alternative output states and their probability amplitudes are represented by the multiple paths.

Mainly because of conservation laws (e.g., energy conservation), the output state of particle1 is correlated with that of particle2, which may be called an entanglement of the two output particles. The functional model of QT measurement accomplishes the entanglement via the formation of common paths for both output particles.
In accordance with the theory, a path that combines the two entangled particles has a single associated probability amplitude. 
The interaction result that contains the multiple paths is called a pw-collection in this paper. The structure of a pw-collection is shown in table 1. The fact that a (single) probability amplitude is associated with a pair of possible exit particle states establishes the entanglements which result from the interaction.

\begin{table}
\caption{\label{label}Structure of pw-collection.}
\begin{tabular} { | c | c | c  | c | }
\hline
paths & pw[1]-state & pw[2]-state  & amplitude  \\

\hline

path-1	  &   pw[1]-state$_{1}  $    &  pw[2]-state$_{1} $  & ampl-1  \\

path-2	 &   pw[1]-state$_{2}  $    &  pw[2]-state$_{2} $   & ampl-2  \\

...	            & ...         & ...    & ...  \\

path-N	 &  pw[1]-state$_{N} $    &  pw[2]-state$_{N} $  & ampl-N  \\
\hline
\end{tabular}
\end{table}  

\subsubsection{Collapse of the wave function}
After the single paths have been selected from the entry particle/waves and an exit particle/wave collection is generated,  entry particle/wave collections are obsolete. The collapse of the wave function means that the pw-collections for the entry particles/waves are discarded. The destruction of the entry pw-collection also affects the particle/waves that are entangled through the entry pw-collection, i.e., the activation of entanglement.

\subsection{From Interactions to Measurement}

Typically, a measurement includes more than one interaction.  At least one interaction has to occur between the measured object and the measurement apparatus. Measurements with multiple interactions between the measured object and the measurement apparatus may also be useful (e.g., a cloud chamber).
Further interactions are usually required within the measurement apparatus to communicate the measured value to an observer.

The need for multiple interactions is a consequence of the fact that, when a specific observable $ S_{meas} $ (e.g., $ S_{meas} = spin $) is to be measured, there are typically only a limited number of interaction types that are capable of producing an interaction output $ S_{out.1} $ (e.g., $ S_{out.1} = position $), which can be interpreted as a mapping of the value of $ S_{meas} $. Even if there exists an interaction type that maps $ S_{meas} $ to $ S_{out.1} $, it is possible that the observable $ S_{out.1} $ cannot directly be observed by an observer, which may require another interaction that maps $ S_{out.1} $ to $ S_{out.2} $, which then may be directly observable.

It is the task of the physicist to design the measurement process, including the series of interactions, in such a way that the actual value of the observable $ S_{meas} $ is determined and properly communicated to the observer. 
As described in section 3.3, the types of interactions that can be used are limited by the laws of QFT that concern interactions. 

An example of a measurement that requires multiple interactions is the Stern-Gerlach experiment for measuring the spin of an electron. After the first interaction between the measured electron and a magnetic field, the spin orientation is reduced (i.e., projected) to a definite value. However, it takes a second interaction to map this measured value to the position of the electron on a screen to make it observable.

\section{The Measurement Problem in Relation to the Proposed Process Model}

In section 2, the measurement problem is described, leading to a set of questions that are either lacking an answer or where the answers provided by the various interpretations of QT are controversial. In the following sections, the questions listed in section 2 are answered from the point of view of the proposed model of the measurement process.
\subsection{Reduction}
\begin{itemize}
\item Does measurement determine the state of a QT object?
\\
In general, measurements have the objective of determining the specific components of the state of an object.
Because QT measurement can only be performed via interactions (and these interactions have to adhere to the laws of QFT), the interactions (and, thus, the measurements) are limited in their ability to communicate the state of the measured object. QT measurement can only determine certain subsets of the state (that existed prior to the measurement). The subset of the state that can be communicated may be interpreted as an imprecise value in some cases.
\item Is the wave function complete?
\\
For the proposed model, the wave function is complete, because it is sufficient to construct the  model of QT interactions and QT measurements.

\item Does measurement always have determinate outcome?
\\
The laws of QFT demand that interactions, including interactions used in measurements,  result in a range of possible outcomes with different probability amplitudes (in section 3.3 "Information Exchange" this is called "probabilistic projection"). Only for exceptional cases such a probabilistic projection represents a determinate outcome.

\item Why is it impossible to measure the non-definite (probability amplitude) values?
\\
A complete non-definite value (i.e., the value of an observable that has multiple values with non-zero probability amplitude) cannot be determined using a measurement for two reasons:
\begin{itemize}
\item An interaction (according to the model of the measurement process) always starts with the selection of single paths for the interacting QT objects. Thus, if the non-definite value is the result of the superposition of multiple paths, this information is lost. 
\item The communication of an exact value would require a "copying" of this value from the measured QT object to one of the output QT objects. The QFT laws for interacting QT particles do not support such a copying process but only
"probabilistic projections" (see section 3.3 "Information Exchange").
\end{itemize}

\item Why can measurements only communicate projected values?
\\
As described in section 3.3, the result of an interaction is always a function of both interacting QT objects. Only in rare special cases can the impact of the MA-object on the measurement value be neglected.

\item Why can certain observables not be measured concurrently?
\\
For a given pair of observables, it may be possible to measure them concurrently. However, due to the laws of QT, it may be impossible that both observables have definite values concurrently. Thus, both observables can be measured, but at least one of the values would be non-definite and can therefore only be measured to a limited extent.

\item When (in the sequence of process steps and under which circumstances) does the transition from probabilities to facts occur? 
\\
Within the model of the measurement process three decision points represent the
transition from probabilities to facts: (1) at the beginning of an interaction the position of the interaction is determined, (2) the selection of a single path from the set of possible multiple paths, (3) the decision for the exiting QT objects.
Item (2) is the basis for the collapse of the wave function. Item (1) and item (2) are key for the explanation of the measurement process.
\end{itemize}

\subsection{Collapse of the wave function}
\begin{itemize}
\item Does the wave function evolve according to the Schr\"odinger equation?
\\
The Schr\"odinger equation (or similar equations like the Dirac equation) describes the evolution of the probability amplitudes (i.e. the wave function) towards possible measurement results. The validity scope of the Schr\"odinger equation, thus ends where the probabilities are transformed into facts. According to the model of the measurement process the transition from probabilities to facts occurs with normal interactions. Interactions are described by the equations of QFT.
\footnote{One might envisage to derive the Schr\"odinger equation from QFT.}
A more complete understanding on how the transition from probabilities to facts occurs, can only be reached by a process model of QFT interactions.
\item What are the criteria for the distinction between interactions that include a collapse of the wave function from those that do not?
\\
Unfortunately, there does not seem to exist any experimental data on this question. From experience (without explicit experiments on this subject), it seems that interactions in quantum optics (i.e., involving photons) mostly do not imply a collapse of the wave function, which suggests that the criteria for the occurrence of the wave function collapse might be that the energy (including the mass) of one of the interacting QT objects has to be much higher than that of the other QT object. 
If one of the interacting QT objects is a bound system (such as an atom, a nucleus, or hadron), the bound system as a whole would have to be taken as one of the interaction objects, except if the other (smaller) QT object has a relatively high energy.
\item Is measurement always coupled with the collapse of the wave function?
\\
Yes, according to this process model.
\item Does the principle of parsimony favour no-collapse theories?
\\
Obviously, this depends on the details of the concrete no-collapse theory. Assuming no collapse at all, but instead a branching into many worlds, is not considered a reasonable alternative because it would not simplify the model at all. The major problems remaining / occurring would be
\begin{enumerate}
\item A criteria and mechanism which distinguishes interactions which imply a collapse (substitute) from those which don't.
\item Non-collapse cannot simply mean letting the incoming paths of the wave functions continue, but would require (a) the creation of multiple new universes for the alternative paths, and (b) copying the complete state of the original universe to the new ones.
\end{enumerate}  
\end{itemize}
\subsection{Transition from Probabilities to Facts}
\begin{itemize}
\item Is the transition from probabilities to facts related exclusively to measurements?
\\
The transition from probabilities to facts is connected to interaction. Interactions, however, do not exclusively occur with measurements.
\item Is the measurement process random or deterministic?
\\
The transition from probabilities to facts (which occurs with an interaction) is a random process step. 
\item What type of interaction constitutes a measurement?
\\
Within this process model, a measurement requires at least one interaction 
which implies an information exchange between the measured QT object and a MA-object of the measurement apparatus.
As described in section 3. QT/QFT knows also interactions where no information is exchanged and which therefore are not suitable for measurements. 
\item When (in the sequence of process steps and under which circumstances) does the transition from probabilities to facts occur? 
\\
See above.
\end{itemize} 

\subsection{Entanglement}

\begin{itemize}
\item When (under which circumstances) is entanglement terminated ?
\\
Within this process model, the collapse of the wave function also terminates entanglements.
\item Does the termination of entanglement always imply modifications of the entangled quantum objects?
\\
Yes, in this process model, the transition from probabilities to facts which occurs with interactions  (1) terminates entanglement and (2) modifies both the entangled QT objects through the elimination of non-selected paths. 
\item Does the modification of an entangled quantum object always terminate the entanglement?
\\
Modifications due to interactions (even if not part of a measurement) always terminate entanglement.
No experimental data are known to the author as to whether interactions which do not result in any modifications terminate entanglements.
\item Is it possible to have a hierarchy of entanglements?
\\
The model of the measurement process does not support hierarchies of entanglement. Support for hierarchical entanglement would be possible but would make the model more complicated.
\item Is entanglement always combined with superposition?
\\
Superposition means multiple paths. Within the model of the measurement process, entanglement can only be realized through multiple paths.
\item Does the measurement process violate locality?
\\
Yes, if it applies to entangled QT objects.
\\
Within the process model, this violation is reflected in the fact that the location where the information that describes the entanglement, i.e., the pw-collection, is undefined.
\end{itemize}
\subsection{Lack of a Functional Description}
\begin{itemize}
\item Does this model of the measurement process provide possible solutions to the measurement problem?
\\
Yes, the solutions and answers are listed above.

\item To what extent can this process model be verified?
\\
Theories on QT measurement are difficult to verify because verification normally requires measurements, the subject to be verified. The verification of the model of the measurement process is less difficult because
\begin{enumerate}
\item The model does not assume any special properties of measurements. The model assumes normal interactions which behave according to the laws of QFT.
\item The model does not make any assumption outside the scope of falsifiable physics.
\item Functional models, in general, can be easily tested using computer simulations. Computer simulations, of course, can only verify that the results of the simulated process are in accordance with the experiments and predictions of QT, in general.
\end{enumerate}
\end{itemize} 

\section{Conclusions}
 
The development of the model of the measurement process originated from the author’s claim that a satisfactory solution to the questions that are usually associated with the measurement problem can only be found if the process of interactions in QT is understood (or at least considered).
Based on the assumptions that the interactions between a measured QT object and the measurement apparatus that occur during a measurement are "normal" interactions (as defined in quantum field theory), a model for the interaction process is described. The major constituents of the model are a set of actions and the overall process within which the actions are embedded.
The model provides answers to questions that are usually associated with the measurement problem (see section 2 and section 4).
The answers given by the process model derive from the following assumptions and claims:
\begin{enumerate}
\item The evolution of the wave function during a measurement process is not just a normal linear progression but a more complicated process which includes the transition from probabilities to facts. (This transition has to become central to QT, rather than being staged at the boundary of QT.)
\item Measurements require interactions between the measured QT object and part of the measurement apparatus; in general, interactions (a) imply transitions from probabilities to facts and (b) have to adhere to the rules and equations of quantum field theory (QFT).
\item Interactions, in general, support only a limited exchange of information between the entering QT objects and the exiting QT objects. This limited exchange of information is called by the author the "probabilistic projection" and is the cause of some of the peculiarities of QT measurements.
\end{enumerate}

The benefits of the process model (as opposed to most other interpretations of QT) is that the model does not require a modification of the equations that define the possible results of a measurement of QT observables (e.g., the Schrödinger equation) nor does it require assumptions outside the scope of falsifiable physics.

\end{document}